\documentclass[%
 reprint,
superscriptaddress,
nofootinbib,
 amsmath,amssymb,
 aps,
 showkeys,
]{revtex4-2}
\usepackage{graphicx}
\usepackage{dcolumn}
\usepackage{bm}
\usepackage{hyperref}
\usepackage[bottom]{footmisc}
\usepackage{footnote}
\usepackage{xcolor}
\usepackage[mathlines]{lineno}
\usepackage{fontenc}
\usepackage{graphicx}
\usepackage{dcolumn}
\usepackage{bm}
\usepackage{booktabs}
\usepackage{array,multirow}
\usepackage{amsmath}
\usepackage{amsfonts}
\usepackage{amssymb}
\usepackage{mathrsfs}
\usepackage{enumerate}
\usepackage{fancyhdr}
\usepackage{xcolor}
\usepackage{graphicx}
\usepackage{listings}
\usepackage{float}
\usepackage{verbatim}
\usepackage{braket}
\usepackage{centernot}
\usepackage{setspace}
\usepackage{endnotes}
\usepackage[normalem]{ulem}
\usepackage{wrapfig}
\usepackage{multirow}
\usepackage{comment}
\usepackage{kantlipsum}
\allowdisplaybreaks
\usepackage{lipsum, babel}
\hypersetup{breaklinks = true, colorlinks = true, citecolor = blue, linkcolor = blue, urlcolor = blue}
\usepackage[mathlines]{lineno}

\newcommand{\be}{\begin{equation}}
\newcommand{\ee}{\end{equation}}
\newcommand{\bea}{\begin{eqnarray}}
\newcommand{\eea}{\end{eqnarray}}

\newcommand{\xpom}{x_\mathbb{P}}

\begin{document}

\title{Probing the onset of maximal entanglement inside the proton in diffractive DIS}
%
%
%
%
\author{Martin~Hentschinski}
\email{martin.hentschinski@udlap.mx}
\affiliation{Departamento de Actuaria, F\'isica y Matem\'aticas, Universidad de las Am\'ericas Puebla, San Andres
Cholula, 72820 Puebla, Mexico}

\author{Dmitri E.~Kharzeev}
\email{Dmitri.Kharzeev@stonybrook.edu}
\affiliation{Center for Nuclear Theory, Department of Physics and Astronomy, Stony Brook University, New York 11794-3800, USA}
\affiliation{Department of Physics, Brookhaven National Laboratory, Upton, New York 11973-5000, USA}

\author{Krzysztof~Kutak}
\email{krzysztof.kutak@ifj.edu.pl}
\affiliation{Institute of Nuclear Physics, Polish Academy of Sciences, ul. Radzikowskiego 152, 31-342,
Krak\'ow, Poland}

\author{Zhoudunming~Tu}
\email{zhoudunming@bnl.gov}
\affiliation{Department of Physics, Brookhaven National Laboratory, Upton, New York 11973, USA}

\date{\today}
\begin{abstract} 
It has been proposed that at small Bjorken $x$, or equivalently at high energy, hadrons represent maximally entangled states of quarks and gluons. This conjecture is in accord with experimental data from the electron-proton collider HERA at the smallest accessible $x$. In this Letter, we propose to study the onset of the maximal entanglement inside the proton using Diffractive Deep Inelastic Scattering. It is shown that the data collected by the H1 Collaboration at HERA allows to probe the transition to the maximal entanglement regime. By relating the entanglement entropy to the entropy of final state hadrons, we find a good agreement with the H1 data using both the exact entropy formula as well as its asymptotic expansion which indicates the presence of a nearly maximally-entangled state.   Finally, future opportunities at the Electron Ion Collider are discussed.
\end{abstract}

\maketitle
 
\section{Introduction}
At the heart of the theory of strong interactions, Quantum Chromodynamics (QCD), there is the phenomenon of \textit{color confinement} that we still do not understand. We know perfectly well that it exists and our own existence is the proof, but its mechanism has been one of the most important unsolved problems in modern physics~\cite{Accardi:2012qut}. Recent advances in quantum information science have allowed to look at this problem from a different perspective~\cite{Klebanov:2007ws,Kharzeev:2017qzs,Beane:2018oxh,Mueller:2019qqj,Lamm:2019uyc,Chakraborty:2020uhf,Liu:2020eoa,Briceno:2020rar,Davoudi:2020yln,deJong:2021wsd,Barata:2021yri,Li:2021kcs,Gong:2021bcp}, see Ref.~\cite{Beck:2023xhh} for a recent review. In fact, confinement can be viewed as an ultimate limit of entanglement as the 
quarks and gluons are not just correlated, but simply cannot exist in isolation. 

In Quantum Mechanics, an isolated proton is a pure quantum state with zero von Neumann entropy. However, when viewed as a collection of quasi-free partons such as in the parton model~\cite{Bjorken:1969ja,Feynman:1969wa,Gribov:1967vfb}, the proton possesses a non-zero entropy associated with different ways to distribute partons in the phase space. To resolve this paradox, a proposal has been made in Ref.~\cite{Kharzeev:2017qzs} that a Deep Inelastic Scattering (DIS) process probes only a part of the wave function of the proton, thus described by a reduced density matrix where the unobserved part is traced over. There is an entanglement entropy associated with the measured reduced density matrix, which represents the entropy associated with the parton distributions. Thus, the DIS process can be viewed as a sudden quench 
of the entangled quantum state of the proton, as a result of which a finite entropy is produced \cite{Kharzeev:2017qzs,Tu:2019ouv,Zhang:2021hra,Kharzeev:2021nzh}. This final state entropy can be measured from the multiplicity distribution of the produced hadrons.
 
More specifically, in an electron-proton ($ep$) DIS process, the virtual photon emitted by the electron has a four-momentum $q$ that probes only a part of the proton wave function with a transverse spatial size of $\sim 1/Q$, where $Q^{2} =-q^2$ characterizes the resolution of the probe. This measurement provides access to a subset of the total density matrix, $\rho$, of the proton. This lack of information about the rest of the proton gives rise to the entanglement entropy, $S_E = - \rm{tr} \rho_A \ln \rho_A$, where the reduced density matrix $\rho_A = \rm{tr}_B \rho$ is obtained by tracing over the unobserved degrees of freedom of the total density matrix $\rho$.  The entropy production in high energy scattering has been investigated also in Refs.~\cite{Kutak:2011rb,Peschanski:2012cw,Stoffers:2012mn,Kovner:2015hga,Berges:2017hne,Kovner:2018rbf,Armesto:2019mna,Duan:2020jkz,Dvali:2021ooc,Hagiwara:2017uaz,Neill:2018uqw,Liu:2022ohy,Liu:2022hto,Dumitru:2022tud,Dumitru:2023fih,Ehlers:2022oal,Duan:2023zls,Liu:2023eve,Asadi:2023bat,Liu:2022bru,Kou:2022dkw,Dumitru:2023qee}. 
 
Based on an explicit model of QCD evolution at small Bjorken $x$, it has been conjectured~\cite{Kharzeev:2017qzs} that the inclusive DIS process probes a proton in the maximally entangled state, i.e., in a state where a large number of partonic micro-states occur with equal probabilities, $P_n(Y) = 1/\langle n\rangle$. Here $Y$ is the rapidity and $n$ is the number of resolved constituents of the proton. This maximally entangled state corresponds to an entropy $S=\ln n$, which has been confirmed by comparison of calculations to data both in proton-proton collisions~\cite{Tu:2019ouv} and inclusive $ep$ DIS~\cite{Hentschinski:2022rsa,Kharzeev:2021yyf,Hentschinski:2021aux,H1:2020zpd}.
Therefore, the questions of interest that arises from these findings are, how does the maximally entangled state emerge, and whether there are conditions under which the constituents of the proton are \textit{not} maximally entangled? 

It has been found that in non-diffractive DIS process at sufficiently small $x$, one probes a maximally entangled state of the proton~\cite{Hentschinski:2021aux,Kharzeev:2021yyf}. 
However, it is known that $\sim 15\%$~\cite{Wolf:2009jm,Armesto:2019gxy} of the inclusive DIS cross section measured at HERA is from diffractive processes, where a rapidity gap in the distribution of the hadronic final states is observed (for a review see~\cite{Frankfurt:2022jns}). These diffractive processes are believed to probe 
different components of the parton wave function of the proton, in which the parton evolution is ``delayed'' by the presence of the rapidity gap\footnote{For other work on the relation of diffraction at high energy scattering and entanglement we refer the Reader to \cite{Peschanski:2019yah}} \cite{Kovchegov:1999ji, Hentschinski:2005er, Kozlov:2006cu, Mueller:2018ned, Le:2021qwx}.
%

In this Letter, we present the first study of the entanglement entropy associated with diffractive deep elastic scattering (DDIS) processes, based on a dipole cascade model \cite{Mueller:1994gb}. To validate our model, we compare it to the  published data from the H1 Collaboration on charged particle multiplicity distributions in DDIS at the top HERA energy. Finally, we discuss future opportunities at the upcoming Electron-Ion Collider (EIC) at Brookhaven National Laboratory.

\section{Cascade model for diffraction}

\begin{figure}
     \centering
     \includegraphics[width=.49\textwidth]{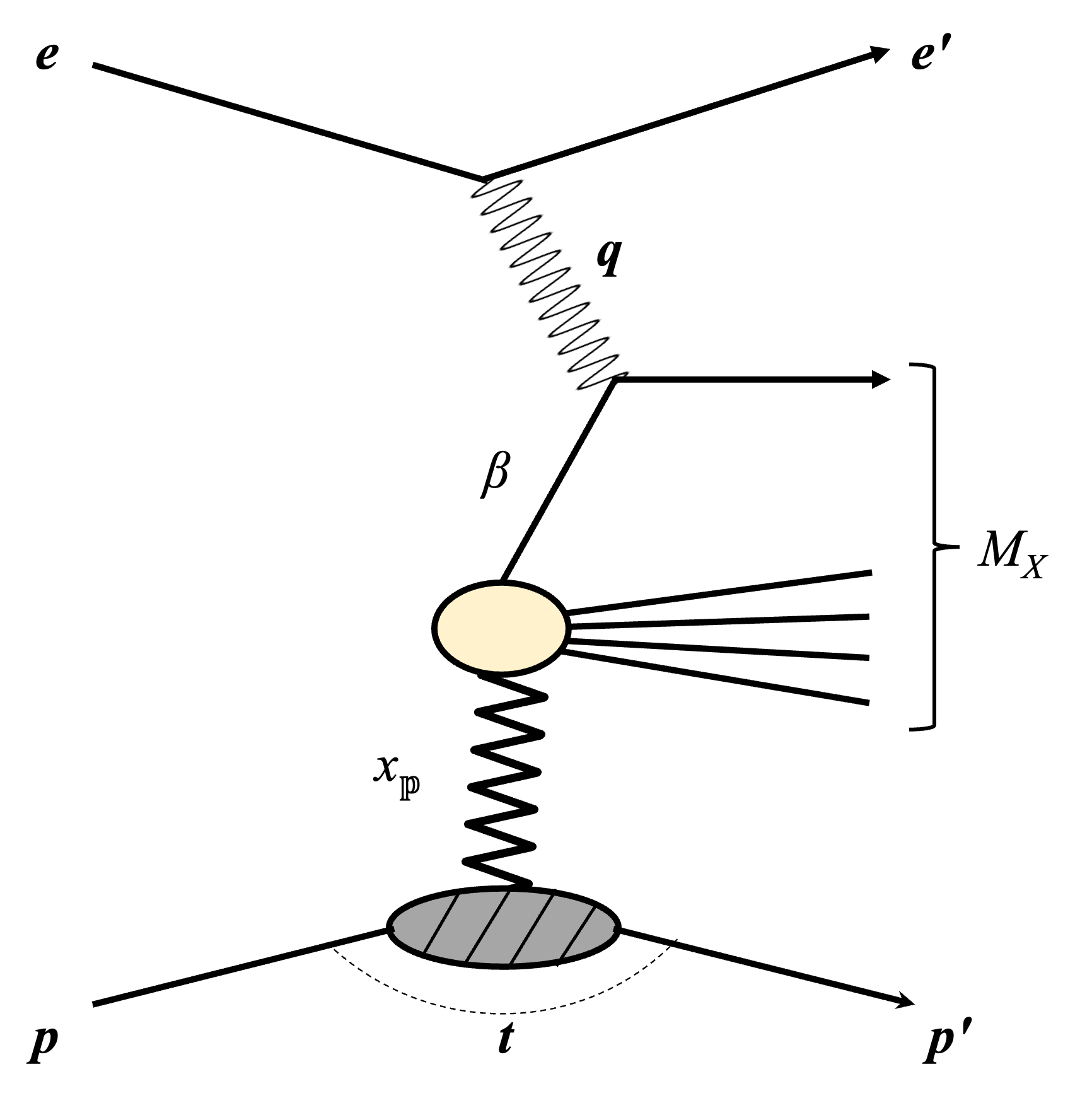}
     \caption{Kinematics of the neutral current diffractive DIS process $ep\rightarrow ep X$.}
     \label{fig:kinematics}
 \end{figure}

We consider DDIS of an electron on a  proton target. As for inclusive DIS events, these events are characterized by the virtuality of the photon $q^2 = -Q^2$ as well as Bjorken $x = Q^2/2 p\cdot q$, where $q$ and $p$ denote the four-momentum of the virtual photon and proton respectively, see also Fig.~\ref{fig:kinematics}. Diffractive events are further characterized by $\xpom$ which denotes the proton's momentum fraction carried by the Pomeron. The magnitude of the rapidity gap  $y_0$ is related to $\xpom$ by $y_0 \simeq \ln 1/\xpom$. The variable $\beta$ denotes the Pomeron's momentum fraction carried by the quark interacting with the virtual photon. For collinear kinematics, $x =  \beta \cdot \xpom$. With $Y = \ln 1/x$, the width of the rapidity interval occupied by the diffractive system $X$ formed in the collision is $y_X = Y - y_0 \simeq \ln 1/\beta$. 

For large invariant mass $M_X$ or small values of $\beta$ of the diffractive system $X$, using factorization and the limit of large number of colors, the diffractive system can be described as a set of color dipoles~\cite{Kovchegov:1999ji, Hentschinski:2005er, Kozlov:2006cu, Mueller:2018ned, Le:2021qwx}. Within the 1+1 dimensional model for the distribution of dipoles \cite{Mueller:1994gb} used in \cite{Levin:2003nc,Kharzeev:2017qzs}, the probability $p^D_n(y_X)$ to have exactly $n$ dipoles is described by the following cascade equation:
\begin{equation}
    \label{eq:cascade}
   \frac{\partial p_n^D(y_X)}{\partial y_X} =-n\Delta  p_n^D ( y_X)+(n-1)\Delta p_{n-1}^D( y_X),
\end{equation}
where $\Delta$ controls the rate at which the number of dipoles grows. 
In the following we consider a slight generalization of the solution to this equation used for the inclusive case,  i.e,
\begin{align}
     \label{eq:gen_solution}
     p_n^D(y_X) & = \frac{1}{C} e^{-\Delta y_X} \left(
     1- \frac{1}{C} e^{-\Delta y_X}
     \right)^{n-1}.
 \end{align}
Introducing the additional  constant $C \geq 1$ allows to take into account the possibility that more than one dipole exists at $y_X=0$. For diffractive reactions, the exchanged Pomeron serves as a source for the generation of diffractive dipoles and therefore $p_{n\geq 1} (0) \neq 0$ is possible, see also \cite{Kovchegov:1999ji, Hentschinski:2005er, Kozlov:2006cu, Mueller:2018ned, Le:2021qwx}. With the above modification we have for the average number of dipoles,
\begin{align}
    \label{eq:partonNumber}
    \left\langle \frac{dn(\beta)}{d \ln 1/\beta} \right \rangle & = \sum_n n p^D_n (y_X)= C\left(\frac{1}{\beta}\right)^\Delta,
\end{align}
which can be identified with the number of partons per unit of $\ln 1/\beta$. The latter can be related to the diffractive parton distribution functions (PDF) $\beta \xpom f(\beta, \xpom)$ in the low $\beta$ region. 

\section{Diffractive DIS data}
Data used in this Letter was collected by the H1 Collaboration~\cite{H1:1998xpp} during the HERA 1 period. The measurements of charged particle multiplicity distributions were performed in the rest frame of the hadronic final-state $X$.  A  minimum pseudo-rapidity gap of $\sim 4.3$ units was imposed. The data analysis was done separately for the forward and backward hemispheres. To evaluate the entanglement entropy, one should include all charged particles in the diffractive final states. Therefore, we combine the measured multiplicity distributions from forward and backward hemispheres into a total charged particle multiplicity distribution, which serves as the input to calculate the hadron entropy. The detailed procedure of obtaining the total charged particle multiplicity distributions are provided in the Supplemental Material, which includes Ref.~\cite{H1:1996ovs}.  

The entropy of the final hadronic state is calculated as follows:
\be\label{eq:entropy}
S_{\text{hadron}} = -\sum{P_{n}\log{P_{n}}},
\ee
\noindent where $P_{n}$ is the probability to detect $n$ charged hadrons. Similar analyses were done in Refs.~\cite{Tu:2019ouv, H1:2020zpd}. 
Comparing to the inclusive DIS measurement of hadron entropy in Ref.~\cite{H1:2020zpd}, the covered phase space of charged hadrons in the DDIS measurement was larger~\cite{H1:1998xpp}, with track selection within $|\eta_{\rm lab}|<2.0$ and transverse momentum larger than 100 $\rm MeV/c$. 


 \section{Numerical results\\ and the model comparison}
 In the following, we compare our model to the data from the H1 Collaboration~\cite{H1:1998xpp}. 
 We use a description based on a direct extrapolation of the 1+1 dimensional model Eqs.~\eqref{eq:cascade} and~\eqref{eq:gen_solution} to the relevant values of $\beta$. To this end, we use the fact that the number of partons and  the number of dipoles coincide in the low $\beta$ region  and are directly given by the corresponding leading order diffractive PDFs. To compare with the available data set, we average over $Q^2$ and  integrate over the region probed in $\xpom$:
\begin{align}
\label{eq:dip}
\left\langle \frac{dn (\beta)}{d\ln 1/\beta}\right\rangle  & =  \frac{1}{Q^2_{\text{max}} - Q^2_{\text{min}}}
 \int\limits_{Q^2_{\text{min}}}^{Q^2_{max}} dQ^2
 \int \limits^{x_{\mathbb{P},\text{max}}}_{x_{\mathbb{P},\text{min}}} d\xpom  \notag\\ 
 & \beta\left[
  f_{\Sigma/p}^D \left(\beta, \xpom, Q^2 \right) 
 + f_{g/p}^D \left(\beta, \xpom, Q^2 \right) 
 \right],
\end{align}
 where
 \begin{align}
    f_{\Sigma/p}^D  \left(\beta, \xpom, Q^2 \right)  &= \sum_{f=1}^{n_f}
    \left[ f^D_{q_f/p}\left(\beta, \xpom, Q^2 \right)  \right.\notag \\&
   \left.  \qquad + f^D_{\bar{q}_f/p}\left(\beta, \xpom, Q^2 \right) \right],
 \end{align}
 and $Q_{\text{min}}^2 = 7.5$~GeV$^2$, $Q_{\text{max}}^2 = 100$~GeV$^2$, while $x_{\mathbb{P},\text{min}} = 0.0003$ and $x_{\mathbb{P},\text{max}} = 0.05$. The selected phase space is chosen to reproduce the phase space in which the H1 data was analyzed.
 
 To fix the free parameters of the model, $C$ and $\Delta$, we impose that the average number of dipoles given by Eq.(\ref{eq:partonNumber}) in the low $\beta$ region, $\beta  \in [10^{-5}, 10^{-3}]$ should agree with predictions based on diffractive PDFs, for which we use the leading order results GKG18-DPDFs (Set A),  provided by the authors of \cite{Goharipour:2018yov}. In particular we use for the diffractive PDF of the parton $i$ the following parametrization
 \begin{align}
    \beta  f^D_{i/p}(\beta, \xpom, Q^2) 
    & = F_{\mathbb{P}/p}(\xpom) 
    \cdot 
   \beta f_{i/\mathbb{P}}(\beta, Q^2),
 \end{align}
 with  Pomeron flux factor
 \begin{align}
     F_{\mathbb{P}/p}(\xpom)  & = A_P \cdot \frac{D}{\xpom^{\lambda_P}},
 \end{align}
where $A_P =  2.39187$,  $D=0.142735$ and $\lambda_P = 1.185$. We then find  $\Delta = 0.287 \pm 0.006 \pm 0.050$ and $C = 4.20 \pm 0.29 \pm 0.52$, where the uncertainties are evaluated by varying different $\beta$ ranges (first uncertainty) and PDF (second uncertainty). See the detailed procedure in the Supplemental Material. Invoking parton-hadron duality, this number should approximately agree with the average number of hadrons measured in DDIS. 

As noted in \cite{Hentschinski:2022rsa}, experiments measure only the charged hadron multiplicity and one assumes 
 \begin{align}
     \left\langle \frac{dn (\beta)}{d\beta}\right\rangle_{\text{charged}} \simeq \frac{2}{3} \left\langle \frac{dn (\beta)}{d\beta}\right\rangle.
 \end{align}
 To describe the entropy of charged hadrons, we thus replace in our expression $C \to C' = 2/3 \cdot C = 2.80 \pm 0.19 \pm 0.34$. Using the parameters listed above, the probability distribution in Eq.~\eqref{eq:gen_solution} yields the average number of partons. This can be used to estimate the number of charged hadrons, if the ratio between charged and neutral particle yield is taken to be a constant, {\it i.e.,} independent of $\beta$. 
 \begin{figure}
 \includegraphics[width=.49\textwidth]{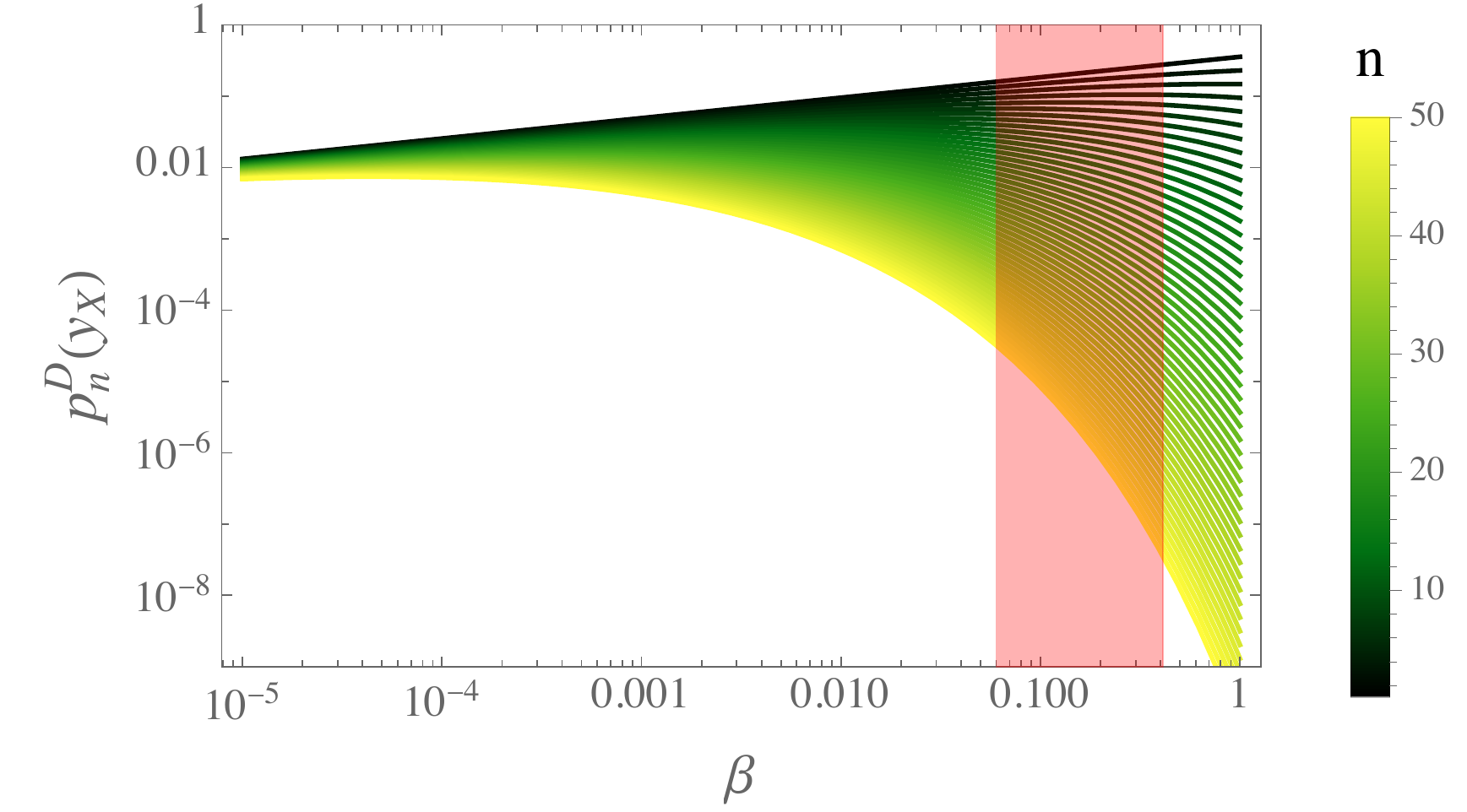}
     \caption{Probabilities $p_n(y_X)$ with $y_X = \ln(1/\beta)$ as extracted from leading order diffractive PDFs for $n=1, \ldots, 50$ for the charged hadron multiplicities. The shaded region indicates the region in $\beta$ probed by the H1 data set.}
     \label{fig:pn}
 \end{figure}

 The resulting probability distribution is illustrated in Fig.~\ref{fig:pn} for $n = 1, \ldots, 50$. In the low $\beta$ region,  the probabilities $p^D_n$ become equal. In the limit $\beta \to 0$ the probability distribution is therefore constant (different multiplicities have equal probabilities) and the entropy reaches a maximum, corresponding to a maximally entangled state. At moderate values of $\beta \in [0.06, 0.41]$, probed by the currently available data set, we observe the gradual transition to the maximally entangled regime. 
Away from the maximally entangled region,   configurations with a few partons have a considerably higher probability than those with many partons -- therefore the entropy does not reach its maximal value.

To compare with hadron entropy extracted from the H1 charged hadron multiplicity distribution, 
we assume (in accord with the local parton-hadron duality~\cite{Dokshitzer:1991eq}) that the multiplicity distributions of hadrons and dipoles are the same, $p_N = p^D_n$. We thus  
use the expression for the hadron entropy (\ref{eq:entropy}) with the dipole probabilities given by Eq.(\ref{eq:gen_solution}). In the maximally entangled regime, all dipole multiplicity probabilities become equal; we write down this universal value as $p^D_n \equiv 1/Z$. The entanglement entropy then takes the form
 \begin{align}
 \label{eq:entr}
     S(Z) & =  -\sum_n p^D_n \ln p^D_n = (1-Z) \ln \frac{Z-1}{Z} + \ln Z.
 \end{align}
We can perform the Taylor expansion of this formula at $Z\to \infty$, when the number of partonic microstates becomes large. This yields 
 \begin{align}
 \label{eq:entrA}
     S_{\text{asym.}}(Z) & =   \ln Z + 1 + {\cal O}(1/Z),&
 \end{align}
\noindent which describes a maximally entangled state and is only applicable in the low $\beta$ region. In the truly asymptotic region $\beta \to 0$, 
 the unity in (\ref{eq:entrA})  may be neglected. However, when the number of partonic microstates is not too large (the case of DDIS in the H1 kinematics), this constant term is still numerically important. For numerical evaluation we use $Z  = C'\beta^{-\Delta}$.

\begin{figure}
     \includegraphics[width=.49\textwidth]{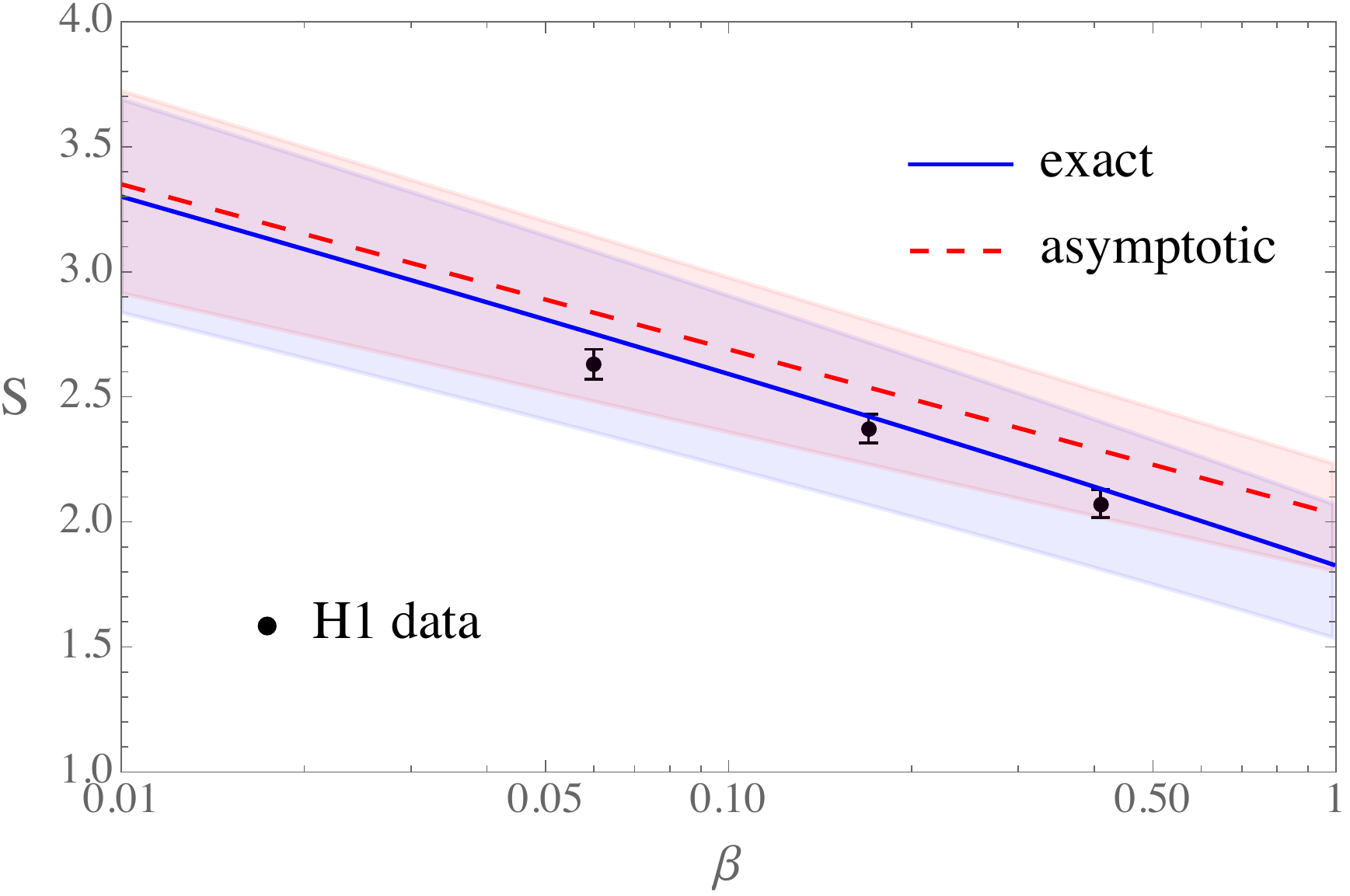}
     \caption{Exact and asymptotic entropy as a function of $\beta$. H1 data~\cite{H1:1998xpp} extracted from the multiplicity distributions are shown, where statistical and systematic uncertainty are added in quadrature and presented as error bars. The theoretical uncertainty bands correspond to PDF and its scale uncertainty added in quadrature, where the scale uncertainty is obtained from the variation of the factorization scale of the leading order diffractive PDFs in the range $Q \to [Q/2, 2Q]$
     }
     \label{fig:entropy_vsdata1}
 \end{figure}

 Our results are shown in Fig.~\ref{fig:entropy_vsdata1} in comparison to the H1 DDIS data.
Uncertainties have been estimated through i) uncertainty from the PDF provided by the LHAPDF~\cite{Buckley:2014ana}, a variation of the factorization scale of the diffractive leading order PDFs in the range $\mu \to [Q/2, 2Q]$, and the model uncertainty on the range of $\beta$ for obtaining parameters $C'$ and $\Delta$. See result with only the PDF uncertainty in the Supplemental Material.
The plot shows that the central value of the result \eqref{eq:entr} is closer to  the data than the asymptotic result \eqref{eq:entrA}, where the goodness-of-fit $\chi^{2}/ndf$ are 2.97 and 17.13, respectively. We see, however, that the curves approach each other at smaller values of $\beta$ indicating that the entanglement entropy reaches its maximal value.

There are a few lessons learned from this study. First, we do not think the quantitative description of our diffractive entanglement entropy model to the H1 data is a coincidence. Entanglement entropy in the context of Deep Inelastic Scattering was introduced in Ref.~\cite{Kharzeev:2017qzs} in 2017 as a new paradigm to understand the nucleon structure at high energy, especially in the nonperturbative regime of QCD where the picture of quasi-free partons breaks down. 
It allows, at least in principle, to directly relate PDFs and final-state hadron production without the use of fragmentation functions (FFs) or other fragmentation frameworks, such as the Lund string model \cite{Sjostrand:2006za}. 

This feature of the entanglement-based approach is seemingly at odds with what we learned during the past decades about particle production in hard processes based on QCD factorization theorems, which describe hadron distribution in $e^+e^-$, DIS, and $pp$ collisions as follows~\cite{Collins:1989gx,Bjorken:1969ja}:
\bea
 \sigma({e^{+}e^{-}\rightarrow hX})&=&\hat{\sigma}\otimes FF,\\
 \sigma({l^{\pm}N\rightarrow hX})&=&\hat{\sigma}\otimes PDF \otimes FF,\\
 \sigma({p_{1}p_{2}\rightarrow hX})&=&\hat{\sigma}\otimes PDF_{1} \otimes PDF_{2} \otimes FF.
\eea

\noindent Here $\hat{\sigma}$ denotes the microscopic QCD cross section for parton scattering, while the FFs describe the transition of the initially produced partons to hadrons. Based on this framework, it seems implausible to not to consider FFs to describe hadron production in high energy collisions. To describe the production of multiple hadrons,  the standard approach is based on semi-classical models like the Lund fragmentation model~\cite{Sjostrand:2006za}. It is not at all trivial to describe the charged particle multiplicity data from hadron colliders based on this model, at least not without significant tuning~\cite{Skands:2014pea}. Also within this approach, there is no direct relation between the structure function and the measured hadron multiplicity distribution.

In contrast to such a description,
the conjecture proposed in Ref.~\cite{Kharzeev:2017qzs} was experimentally confirmed for the first time in the analysis of the multiplicity data from $pp$ collisions at the LHC~\cite{Tu:2019ouv}. 
To further confirm this picture without the ambiguity of two protons in the initial state, a dedicated reanalysis of H1 data taken at HERA~\cite{H1:2020zpd} was performed. 
The evidence again shows the connection between parton density (quarks and gluons) and the final-state hadrons~\cite{Hentschinski:2021aux,Hentschinski:2022rsa} across a wide range of kinematic phase space. The proton is found to be maximally entangled at the top HERA energy in inclusive DIS. This result has added further confidence that entanglement entropy can be a unique probe to the nucleon structure, such as PDFs, and the observed agreement with experimental data is unlikely a coincidence. 

Therefore, in the current study of diffractive DIS, the charged hadron production is again found to be well described by the entanglement model without involving any fragmentation. This demonstrates the broad impact of this new paradigm on describing particle production and the nonperturbative QCD dynamics of entropy production in high energy collisions. 

There is another important lesson learned from this study for the future DIS experiments at the EIC. The QCD evolution of parton density in rapidity is delayed in DDIS by the rapidity gap. Therefore, to study effects of the rapidity evolution, it is essential to have a large detector coverage to impose different rapidity gaps. Currently, the detector design of the ePIC experiment at the EIC has coverage up to $\sim$3.5--4 in pseudorapidity in the hadron-going direction, which is  significantly larger than  at HERA. In addition, the forward region at the EIC will also have a large acceptance coverage, which enables further control on the rapidity gap size. Quantitative studies of the onset of maximally entangled regime in the ePIC experiment should be performed in the near future.

\section{Conclusions}
In conclusion, we investigated the onset of maximally entangled regime inside the proton in diffractive deep inelastic scattering. 
Using  diffractive parton distribution functions and a dipole cascade model, we described the hadron entropy measured by the H1 experiment. 

We find that the maximally entangled regime sets in at small values of $\beta$, and that the approach to this regime is controlled by the magnitude of the rapidity gap. This is because the rapidity gap delays the QCD evolution in rapidity, and thus delays the onset of the maximal entanglement by reducing the Hilbert space of partonic states. 

By relating the entanglement entropy to the entropy of final state hadrons, we find a good agreement with the H1 data at small $\beta$ using both the exact entropy formula as well as its asymptotic expansion which indicates the presence of a nearly maximally-entangled state. 

Our study opens new possibilities for the investigation of quantum entanglement inside the proton using diffractive deep inelastic scattering at the Electron Ion Collider.
In a broader context the future research direction is focused on a quantitative understanding of entanglement entropy in hadronic reactions: i) We would like to investigate the hard scale ($Q^{2}$) dependence of entanglement entropy in high-energy collisions. The natural candidates are jet production in the forward direction at the LHC as well as existing HERA data where hadronic entropy was measured in a moving rapidity window \cite{H1:2020zpd}.
ii) Another direction is to apply entanglement entropy to the nonperturbative regime, e.g., understanding charged particle production in the photoproduction limit. 
iii) Moreover, entanglement entropy may be a good probe of the dense systems where nonlinear dynamics may play a role, e.g., gluon saturation~\cite{Kharzeev:2017qzs, Hentschinski:2022rsa}. It is found that the gluon density saturates naturally in the picture of a maximally entangled proton at high energies. However, this connection has not been fully explored. One of the  processes most sensitive to gluon saturation in $ep$ and electron-ion collisions is  diffraction~\cite{Accardi:2012qut}.

\section*{Acknowledgements} 
We thank V.~Guzey and H.~Khanpour for providing their codes for diffractive PDFs and S.~Munier for useful correspondence. M.~Hentschinski acknowledges support by
Consejo Nacional de Ciencia y Tecnolog\'ia grant number A1 S-43940
(CONACYT-SEP Ciencias B\'asicas). The work of D.~Kharzeev was supported by the U.S. Department of Energy, Office of Science, Office of Nuclear Physics, Grants No. DE-FG88ER41450 and DE-SC0012704 and by the U.S. Department of Energy, Office of Science, National Quantum Information Science Research Centers, Co-design Center for Quantum Advantage (C2QA) under Contract No.DE-SC0012704. The work of K.~ Kutak has been partially supported by  European Union’s Horizon 2020  research  and  innovation  program  under  grant  agreement  No.824093 and the The Kosciuszko Foundation for the Academic year 22/23 for the project ``Entropy of dense system of quarks and gluons”. The work of Z.~Tu is supported by the U.S. Department of Energy under Award DE-SC0012704 and the BNL Laboratory Directed Research and Developement (LDRD) 23-050 project.
K.~ Kutak wants to acknowledge the BNL Nuclear Theory Department for hospitality during the period when the project was initiated.

\bibliography{diffraction}
\end{document}